\documentclass[epsfig,graphics,twocolumn,floatfix,mathbbm,prl]{revtex4}
\usepackage{graphicx}
\usepackage{tabularx}
\usepackage{epstopdf}

\begin{document}

\title{Coherent spin control at the quantum level in an ensemble-based optical memory}

\author{Pierre Jobez$^{1}$}
\thanks{These authors contributed equally to this work.}
\author{Cyril Laplane$^{1}$}
\thanks{These authors contributed equally to this work.}
\author{Nuala Timoney$^{1}$}
\author{Nicolas Gisin$^{1}$}
\author{Alban Ferrier$^{2,3}$}
\author{Philippe Goldner$^{2}$}
\author{Mikael Afzelius$^{1}$}
\email{mikael.afzelius@unige.ch}

\address{$^{1}$Group of Applied Physics, University of Geneva, CH-1211 Geneva 4, Switzerland}
\address{$^{2}$PSL Research University, Chimie ParisTech – CNRS, Institut de Recherche de Chimie Paris, 75005 Paris, France}
\address{$^{3}$Sorbonne Universit\'{e}s, UPMC Univ Paris 06, Paris 75005, France}

\date{\today}

\begin{abstract}
Long-lived quantum memories are essential components of a long-standing goal of remote distribution of entanglement in quantum networks. These can be realized by storing the quantum states of light as single-spin excitations in atomic ensembles. However, spin states are often subjected to different dephasing processes that limit the storage time, which in principle could be overcome using spin-echo techniques.  Theoretical studies have suggested this to be challenging due to unavoidable spontaneous emission noise in ensemble-based quantum memories. Here we demonstrate spin-echo manipulation of a mean spin excitation of 1 in a large solid-state ensemble, generated through storage of a weak optical pulse. After a storage time of about 1 ms we optically read out the spin excitation with a high signal-to-noise ratio. Our results pave the way for long-duration optical quantum storage using spin-echo techniques for any ensemble-based memory.
\end{abstract}

\maketitle

Long-distance distribution of entanglement is an outstanding challenge in quantum information science, which would enable long-distance quantum communication, distributed quantum simulations and large-scale quantum networks \cite{Kimble2008}. The distribution of entanglement over large scales ($>$1000 km) using optical fibers requires quantum repeaters \cite{Sangouard2011}, which in turn need quantum memories \cite{Bussieres2013} with long storage times (milliseconds and beyond). Spin-states in atomic ensembles can provide the required long coherence times, both in laser-cooled alkali vapours \cite{Bao2012,Dudin2013} and rare-earth-ion doped crystals \cite{Longdell2005,Lovric2013,Heinze2013}, while also providing strong light-matter coupling through high number densities. However, the spin-wave coherence often dephases due to inhomogeneous spin broadening ($T_2^*$) and/or fluctuations in the surrounding bath ($T_2$). In principle storage times beyond the dephasing time can be achieved using spin echo techniques, which requires manipulating the spins with population-inverting pulses (e.g. $\pi$ pulses). In the limit of spin-bath dephasing ($T_2$ limited), multi-pulse spin echo techniques can actively decouple the spins from the bath, known as dynamical decoupling, where the spin population relaxation time ($T_1$) sets a fundamental limit.

This general approach has been successfully applied in the quantum regime for single qubit systems \cite{Lange2010,Bylander2011,Pla2012,Piltz2014}. For ensemble-based optical memories, however, it has only been applied to storage of bright classical pulses \cite{Longdell2005,Lovric2013,Heinze2013,Dudin2013}. The purpose of our experiment is to investigate if this approach can also be applied to quantum storage. Optical quantum storage results in a single spin-wave excitation delocalized over an ensemble with a macroscopic number of atoms. The challenge is thus to avoid populating the relevant spin state with many spins, due to unavoidable imperfections in the population-inversion pulses. In 2004, Johnsson and M$\o$lmer \cite{Johnsson2004} argued that imperfections in the pulses would cause an intrinsic source of photon noise, making high-fidelity single-photon storage virtually impossible. A precision of the population inversion pulses of $<1/N$ would be required \cite{Johnsson2004}, $N$ being the number of relevant spins ($N$ is 10$^{12}$ in our case). In 2011 Heshami \textit{et al.} \cite{Heshami2011} made a more extensive theoretical study of the applicability of spin-echo manipulation for quantum storage. They showed that the collective enhancement effect at the heart of the light-matter interaction in an ensemble provides a powerful spatial filter, thereby reducing the required precision. We emphasize that coherent storage of bright light pulses, eg. as done in Refs \cite{Longdell2005,Lovric2013,Heinze2013,Dudin2013}, does not allow to address this question.

\begin{figure*}
    \centering
    \includegraphics[width=.95\textwidth]{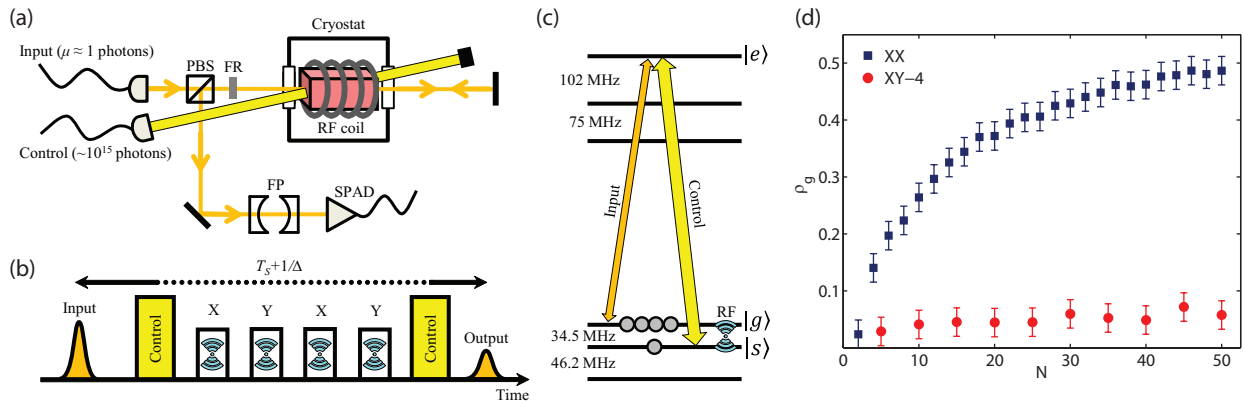}
    \caption{AFC spin-wave storage experiment. In (a) a simplified sketch of the experimental set up is shown. A highly coherent continuous-wave 580.04 nm laser is split into an input mode and a control mode, with about 560 mW of power in the control mode. The output mode crosses the input mode inside the 1 cm long crystal, which is cooled to around 4 K. The input mode is in a double-pass configuration to increase the optical depth. The output mode is sent through a Fabry-Perot (FP) cavity (2.5 MHz bandwidth) to filter out photon emission noise from the crystal at other frequencies (see \cite{Timoney2013}). The single-photon avalanche diode (SPAD) was gated with an AOM not shown. PBS = Polarization Beam Splitter. FR=Faraday Rotator. In (b) the timing of the storage sequence is shown. The total storage time is $1/\Delta+T_S$, where $T_S$ is the spacing between the control pulses. The RF spin-echo sequence is inserted in between the optical control pulses. The RF pulses at 34.5 MHz are applied using a coil placed around the crystal (see (a)). In (c) the hyperfine states of the ground and excited states and the two transitions of the chosen $\Lambda$-system in Eu$^{3+}$:Y$_2$SiO$_5$ are shown. In (d) we show the characterization of the population inversion precision of the XX and XY-4 sequences. Initially all ions are polarized into $|s\rangle$. The relative population in $|g\rangle$ $\rho_g$ is measured after applying $N$ sequences. The error bars represent the error in the absorption measurement used to estimate $\rho_g$. More experimental details are given in the text.}
\end{figure*}

In this work we perform a critical experimental test of the use of spin-echoes in the context of storing light pulses at the single-photon level in an ensemble. To this end we store weak coherent states of light $|\alpha \rangle$ as spin-wave excitations in a Eu$^{3+}$:Y$_2$SiO$_5$ crystal, with mean-photon-numbers $|\alpha|^2=\mu$ in the range of 1 to 2. The extremely weak spin-wave excitation ($\leq 1$) generated through the optical storage is then manipulated by a particular two-axis sequence of population-inversion pulses to reach a spin-wave storage time of about 1 ms. The optical read out of the spin excitation results in a signal-to-noise ratio (SNR) in the range of 5 to 10 for these values of $\mu$. Our results show that the intrinsic noise can be low enough to perform quantum level storage, as predicted by Heshami \textit{et al.}, provided that the spin-echo sequence is tailored specifically to reduce this noise. This paves the way for extremely long duration quantum storage in both laser-cooled gases \cite{Dudin2013} and rare-earth-ion doped crystals \cite{Heinze2013}.

We briefly compare our memory to other reversible (in-out) optical memories working at the single-photon level, in terms of storage time. In an in-out memory the light is written into the memory and subsequently read-out from the memory. Most of such storage experiments reached storage times $\leq$15 $\mu$s \cite{Choi2008, Hosseini2011, Nicolas2014}, with a few notable exceptions. In Ref. \cite{Specht2011} 184 $\mu$s was obtained using a trapped, single $^{87}$Rb atom. The same group also achieved 470 $\mu$s in a BEC of $^{87}$Rb atoms \cite{Lettner2011}. In Ref. \cite{Xu2013} 1.6 ms was reached in a laser-cooled $^{87}$Rb atomic cloud. Our results are thus comparable to the longest storage times for in-out memories. Longer (up to 100 ms) spin storage times were achieved using techniques where the light was either only written into \cite{Julsgaard2004} or read-out from the memory \cite{Bao2012, Radnaev2010}. It is worth noting that an in-out memory is generally less efficient, for the same device, since it is the product of the probabilities of writing into and reading out from the memory. It is thus difficult to compare the performances of memories belonging to different categories. Also, some quantum repeater schemes specifically require in-out memories \cite{Sangouard2011}.

The storage scheme we employ in this demonstration is an atomic frequency comb (AFC) memory with spin-wave storage \cite{Afzelius2009a}. In short it is based on the creation of a frequency grating (the comb) with periodicity $\Delta$ in the absorption profile of an inhomogeneous optical transition. Its interaction with an input pulse leads to an AFC echo after a time $1/\Delta$, such that the comb acts like a variable delay line. The AFC echo scheme has been used in a variety of quantum optics experiments, such as storage of time-energy entanglement \cite{Saglamyurek2011,Clausen2011}, heralded single photons \cite{Clausen2012,Rielander2014} and teleportation from a telecom photon to a memory \cite{Bussieres2014}. To reach longer storage times and on-demand read out, the optical excitation is written to a spin state using an optical population inversion pulse called the control pulse (see Figure 1(b)-(c)). After the spin-wave storage time $T_S$ an identical control pulse re-establishes the optical coherence, which leads to a memory output after a total memory time of $1/\Delta+T_S$.  This full memory scheme, called an AFC spin-wave memory, can in principle perform efficient multi-mode storage for durations only limited by the bath fluctuations ($T_2$ limited), provided that spin-echo techniques are used to compensate for the inhomogeneous spin linewidth.

\begin{figure*}
    \centering
    \includegraphics[width=1\textwidth]{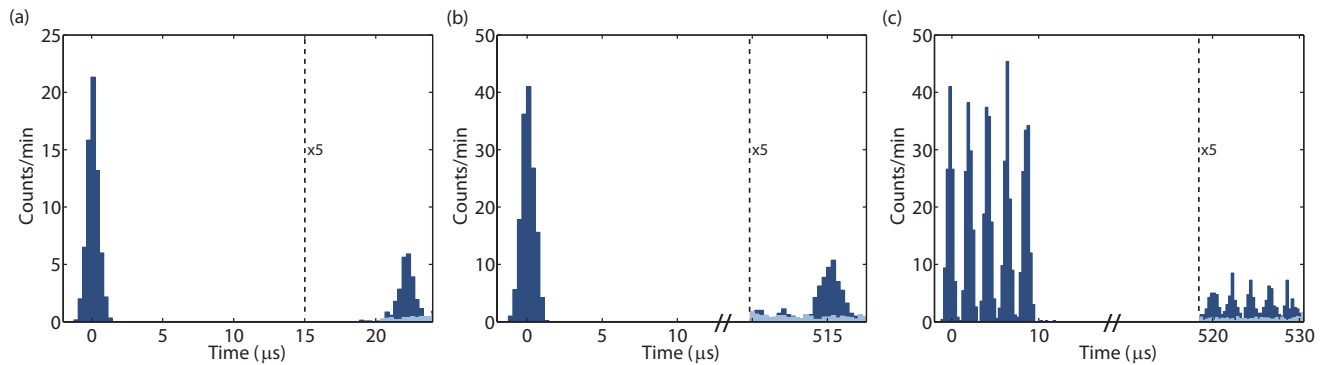}
    \caption{Photon counting histograms. In (a) we show AFC spin-wave storage without spin-echo manipulation. In (b) and (c) we show spin-wave storage with spin-echo manipulation of a single mode (b) and of 5 input modes (c). Each histogram shows data recorded with (dark trace) and without (bright trace) the input pulse, which allows one to measure the SNR in the output mode. The SNR is 11$\pm$2 in (a), 10$\pm$2 in (b) and 7$\pm$1 in (c). The input mode(s) has(ve) a mean photon-number(s) of $\mu$=1.1$\pm$0.1 in (a), $\mu$=2.0$\pm$0.1 in (b) and (c). The storage efficiencies were here (a) $\eta=5.7\pm$0.4 \%, (b) $\eta=5.1\pm$0.4 \% and (c) $\eta=3.1\pm$0.3 \%. In the case of 5-mode storage all parameters are given as averages over the modes. Indicated errors are statistical.}
\end{figure*}

The present AFC spin-wave memory is based on a custom-grown $^{151}$Eu$^{3+}$:Y$_2$SiO$_5$ crystal, with a $^{151}$Eu$^{3+}$ concentration of 1000 ppm. We use the yellow $^7$F$_0$ $\rightarrow$ $^5$D$_0$ transition at 580.04 nm, which has an extremely narrow homogeneous broadening \cite{Koenz2003} and long spin coherence \cite{Arcangeli2014} and population \cite{Koenz2003} lifetimes at cryogenic temperatures. The relevant energy levels and a schematic of the experimental set up is shown in Fig. 1(a)-(c). The isotopically enriched $^{151}$Eu$^{3+}$ doping results in a larger optical depth (absorption coefficient $\alpha$=2.6 cm$^{-1}$) as compared to a natural abundance of Eu$^{3+}$ isotopes. The AFC preparation sequence creates a 2 MHz wide comb on the $|g\rangle$-$|e\rangle$ transition through precise spectral holeburning (see \cite{Jobez2014} for details). It is followed by the storage sequence (see Figure 1(b)) which is repeated 18 times in order to increase the effective rate of the experiment.

The performance of the AFC spin-wave memory is first characterized without applying the spin echo sequence. In this case the storage time is limited by $T_2^* \approx 20$ $\mu$s, corresponding to a spin linewidth of 27 kHz (FWHM). Figure 2(a) shows an example of storage with a mean photon number of $\mu$=1.1$\pm$0.1 in the input mode. The spin-wave storage time was set to $T_S$=11 $\mu$s, resulting in an overall memory efficiency of $\eta=5.7\pm$0.4 \%. The noise of the memory is measured by blocking the input state while executing the complete memory scheme, which gives an unconditional noise probability of $p_n=5\pm1 \cdot 10^{-3}$. Since the overall efficiency is much higher than the noise floor, the memory has a high SNR at the single photon level, as seen in Figure 2(a). The source of the noise is a mixture of incoherent and coherent emissions from the active $^{151}$Eu$^{3+}$ ions, caused by the application of the second optical control pulse. The noise was characterized in more detail in Ref. \cite{Timoney2013}. The level of noise we observe here is similar, but the SNR is one order of magnitude higher which can be attributed to a significantly increased memory efficiency. This is possible due to an optimized AFC preparation \cite{Jobez2014} and a custom-grown Eu$^{3+}$:Y$_2$SiO$_5$ crystal which has been optimized in terms of absorption coefficient and optical inhomogeneous broadening. The higher optical depth of this crystal increases the maximum storage efficiency.

The performance of the memory can also be expressed by $\mu_1$, which we define as the mean photon number in the input that results in a SNR of 1 in the output, i.e. $\mu_1=p_n/\eta$, which is $\mu_1=0.1\pm0.02$ for the memory without spin-echo manipulation. We here consider the theoretical upper limitation of the storage fidelity due to $\mu_1$ in the case of storing a qubit encoded into a true single photon. In this case we can consider the fidelity conditional on the detection of a photon at the output of the memory (post-selected). A straightforward calculation shows that a memory characterized by a certain $\mu_1$, for a single mode, can achieve a two-mode qubit storage with a fidelity given by $F=(1+\mu_1/p)/(1+2\mu_1/p)$, where $p$ is the probability of having the qubit before the memory. This is valid assuming that the noise is state independent (white noise spectrum) and that the fidelity is only limited by noise (complete phase coherence). Storage of a qubit encoded into a true single photon has a classical fidelity limit of $F=2/3$ \cite{Massar1995}, which is surpassed if $p$ exceeds $\mu_1$. Since $p$ is a probability it follows that quantum storage can be achieved if $\mu_1<p<1$. In this regime one can also preserve the non-classical correlations when storing a single photon out of a two-mode squeezed state. Hence we use this parameter to qualify the potential performance of our memory in the quantum regime.

We now turn to the AFC spin-wave storage experiments in the milliseconds regime, where a spin-echo sequence is applied in between the optical control pulses, see Figure 1(b). The main challenge of spin-echo manipulation is to maintain a low unconditional noise probability in order to allow operation in the quantum regime. We therefore first investigate the precision of two different spin-echo sequences in order to minimize the impact of this noise source.

The precision of the spin-echo sequence depends on the precision of the individual population inversion pulses and the design of the sequence itself. Note that one requires a sequence with an even number of pulses to restore the weak spin excitation in $|s\rangle$ before optically reading out the memory. To increase the precision and robustness per pulse one can use chirped adiabatic pulses, which also allow a more uniform manipulation of the spins over the entire spin linewidth. We theoretically estimate the error per pulse to be around 1\%. Two possible spin-echo sequences were tested in terms of population inversion precision. These are the single-axis Carr-Purcell sequence, which we denote XX, and the dual-axis XY-4 sequence. The XY-4 spin-echo sequence \cite{Maudsley1986} consists of four population-inversion pulses, with a periodic spacing of $T_S/4$ between the pulses (see Figure 1(b)). The pulses perform successive $\pi$ rotations around the X and Y axis of the Bloch sphere (XYXY). The more conventional CP sequence performs two rotations around the same X axis. The XY-4 sequence was proposed because it is more robust to errors in the $\pi$ rotations, as compared to the CP sequence \cite{Maudsley1986}. It is also more robust with respect to the phase of the initial spin state \cite{Souza2011}. This is relevant in our experiment since the initial spin state generated through optical storage has a fluctuating phase.

To experimentally characterize the population-inversion precision we first spin polarize all ions into $|s\rangle$ by optical pumping and then apply several XX or XY-4 sequences. The relative spin population $\rho_g$ in $|g\rangle$ is estimated after each sequence by an absorption measurement. Given that each sequence consists of an even number of pulses, unit efficiency would result in no measurable population in state $|g\rangle$. In Fig. 1(d) it is seen that the XY-4 sequence greatly outperforms the XX sequence in terms of population inversion efficiency. The XX sequence results in a complete thermalization of the $|g\rangle$ and $|s\rangle$ states after about 50 sequences, whereas the XY-4 sequence results in only a small fractional population in $|g\rangle$ that is hardly measurable within the error of our absorption measurement. The better performance of XY-4 is attributed to its higher robustness to pulse imperfections \cite{Maudsley1986,Souza2011}. From the data we put an upper bound of the population error per XY-4 sequence of 0.2$\pm$0.1\%, while we estimate the population error per XX sequence to be 3.6$\pm$0.1\%. It should be pointed out that other, more complex pulse sequences could even work better, such as the KDD sequence \cite{Souza2011}, at the expense of a more complex sequence. It would be interesting to investigate how different sequences increase the population, assuming an initial state close to the pole of the Bloch-sphere having a random phase. This particular situation has not yet been well studied.

The XY-4 sequence can now be applied to the spin-wave storage experiment at the single-photon-level, in order to extend the storage time to a millisecond timescale. Figures 2(b)-(c) shows low-noise spin-wave storage with a mean photon number of $\mu$=2.0$\pm$0.1 for a duration of $T_S$=0.5 ms. We measured a conversion efficiency from the optical mode to the spin wave of 50\%, which implies a mean spin-wave excitation of $\mu_S$=1.0$\pm$0.1. The high SNR observed in the output mode clearly underlines the ability to perform precise manipulation of an extremely weak spin excitation in a large ensemble. In addition we show storage of 5 temporal modes, see Figure 2(c), with a high SNR in all output modes. This is possible due to the ability of the AFC scheme to realize scalable temporal multimode storage \cite{Afzelius2009a}. This ability can lead to an important increase in entanglement distribution rates in quantum repeaters \cite{Sangouard2011}.

\begin{table}
	\begin{center}
		\begin{tabular}{l|l|l|l|l}
		  	$T_S$ (ms) & $\eta$ (\%)  & $\mu_1$ & $p_n$ ($10^{-3}$) & SNR  \\ 
		  	\hline
0.25 & 6.5 $\pm$ 0.5 & 0.24 $\pm$ 0.04 & 16 $\pm$ 2 & 8 $\pm$ 2 \\
0.5 & 5.1 $\pm$ 0.4 & 0.2 $\pm$ 0.04 & 10 $\pm$ 2 & 10 $\pm$ 2 \\
0.75 & 3.5 $\pm$ 0.2 & 0.32 $\pm$ 0.05 & 11 $\pm$ 2 & 6 $\pm$ 1 \\
1 & 2.3 $\pm$ 0.2 & 0.3 $\pm$ 0.06 & 7 $\pm$ 1 & 7 $\pm$ 2 \\
1.25 & 1.4 $\pm$ 0.1 & 0.69 $\pm$ 0.12 & 10 $\pm$ 1 & 3 $\pm$ 1 \\
1.5 & 1 $\pm$ 0.1 & 0.96 $\pm$ 0.18 & 9 $\pm$ 1 & 2 $\pm$ 1 \\
	\end{tabular}

	\end{center}
	\caption{The memory efficiency ($\eta$), the $\mu_1$ parameter, the unconditional noise probability ($p_n$) and the SNR measured for a range of spin storage times ($T_S$). The mean input photon number was $\mu$=2.0$\pm$0.1. These data were taken under the same conditions as those in Figure 2(b). The errors are statistical.}
	\label{TAB:mem_perf}
\end{table}

To further investigate the performance of the memory we fully characterize the unconditional noise and the $\mu_1$ parameter for several storage times in the range of 0.25 to 1.5 ms, see Table 1. The noise level remains low for all storage times and the average noise added with respect to the experiment without RF pulses is $p_n^{RF} = 6 \pm 2 \cdot 10^{-3}$. Using a slightly modified version of the model proposed by Heshami et al. \cite{Heshami2011}, which takes into account several experimental limitations to the memory efficiency, we theoretically estimate an unconditional noise floor of $2\pm1 \cdot 10^{-3}$. To calculate this number we assume a population error of 0.2$\pm$0.1\% per XY-4 sequence as measured before. Our data demonstrates that spin echo manipulation can be performed without significantly increasing the optical read-out noise. A $\mu_1$ parameter below 1 also implies that the memory could work in a quantum regime for storage times up to around 1.25 ms. For longer storage times the loss in efficiency decreases the SNR, hence also the $\mu_1$ parameter, although the noise remains constant within the error bars.

In conclusion our experiment demonstrates the feasibility of using spin-echo techniques for extending spin-wave storage times in optical quantum memories based on ensembles of atoms. This will have important consequences for the long-term goal of quantum networks based on both laser-cooled \cite{Bao2012,Dudin2013} and solid-state optical memories \cite{Longdell2005,Heinze2013}, as well as for the current efforts to store microwave quantum states in spin ensembles using hybrid quantum circuits \cite{Grezes2014,Probst2013}. Furthermore, we have demonstrated multi-mode optical storage on a millisecond time scale in a solid-state memory, the longest storage time reported at the single photon level in a multimode in-out memory. Our characterization of the noise of the memory shows that it can in principle store multimode non-classical states of light.

A very recent experiment demonstrated a spin coherence time of 6 hours using a Eu$^{3+}$:Y$_2$SiO$_5$ crystal \cite{Zhong2015}, which opens up a fascinating perspective of unprecedented long storage times for an optical quantum memory. In that work a magnetic field bias was used to induce a magnetic field-insensitive transition, which in our experiment would strongly reduce the efficiency due to a reduction of the participating number of ions. This could be compensated for using a cavity to enhance the effective absorption depth \cite{Sabooni2013,Jobez2014}. A future challenge is to find the experimental conditions where high efficiency optical spin-wave storage and extremely long storage times through dynamical decoupling can be achieved simultaneously. 

The authors thank F\'{e}lix Bussi\`{e}res, Florian Fr\"{o}wis, Emmanuel Zambrini Cruzeiro and Anthony Martin for useful discussions, as well as Raphael Houlmann and Claudio Barreiro for technical support. This work was financially supported by the Swiss National Centres of Competence in Research (NCCR) project Quantum Science Technology (QSIT), by the European projects SIQS (FET Proactive Integrated Project) and CIPRIS (People Programme (Marie Curie Actions) of the European Union Seventh Framework Programme FP7/2007-2013/ under REA Grant No. 287252) and project Idex ANR-10-IDEX-0001-02 PSL*.

\bibliographystyle{unsrt}

\end{document}